\documentclass[aps,showpacs,twocolumn]{revtex4}
\usepackage{amsmath, amssymb}
\usepackage{graphicx}
\usepackage{bbold}
\begin{document}

\title{Finite Isospin Chiral Perturbation Theory in a Magnetic Field}

\date{\today}

\author{Prabal Adhikari}
\email{prabal@umd.edu}
\affiliation{Maryland Center for Fundamental Physics and the Department of Physics, 
University of Maryland, College Park, MD 20742-4111}

\author{Thomas D. Cohen}
\email{cohen@physics.umd.edu}
\affiliation{Maryland Center for Fundamental Physics and the Department of Physics, 
University of Maryland, College Park, MD 20742-4111}

\author{Julia Sakowitz}
\email{jsakowitz@brearley.org}
\affiliation{The Brearley School, 
New York City, NY 10028-7902}


\begin{abstract}
The phase diagram of finite isospin, zero temperature QCD with the pions coupled to photons in a uniform external magnetic field is explored in the low field, small isospin density regime for which chiral perturbation theory is a valid description. For realistic pion masses, the system behaves as a type-II superconductor: a uniform superconducting state is formed at low-enough magnetic fields, a vortex state for intermediate magnetic fields and finally a normal state for large magnetic fields.  In each these phases (including the vortex phase), the $\pi^{0}$ remains uncondensed just as in the zero-external field problem.   The critical magnetic field where the phase transition from the uniform superconducting state to a vortex state occurs was found numerically.
\end{abstract}

\maketitle
\section{Introduction}
QCD in a background magnetic field is of interest due to its potential relevance to heavy ion collisions, neutron stars and perhaps even the early universe. The study of the QCD at zero density demonstrates a rich array of physical phenomena in the presence of a background magnetic field. The system exhibits dimensional reduction~\cite{Shovkovy}, which leads to stronger pairing between quarks and therefore enhances chiral symmetry breaking~\cite{Shushpanov1, Shushpanov2, Werbos1, Werbos2}. Studies have also investigated how an external magnetic field affects the deconfinement transition temperature in QCD; while the magnetic field enhances chiral symmetry breaking, larger temperatures drive the vacuum towards the restoration of chiral symmetry~\cite{Fraga, Agasian}. Furthermore, it has also been suggested that for large enough magnetic fields, the QCD vacuum may exhibit superconductivity due to the condensation of $\rho$ mesons.~\cite{Chernodub}

While QCD at zero density in a magnetic field can be studied by both analytical and lattice methods, a more relevant question in the study of neutron stars relates to how magnetic fields affect finite density matter. This problem is largely unsolved except for a limited regime of asymptotically large baryon densities where a phase of color superconducting is expected to form~\cite{Gorbar}. In particular, the effect of magnetic fields on baryons may significantly affect the properties of magnetars---neutron stars with large magnetic fields~\cite{Harding}. However, the study of finite baryon densities problems using lattice methods are hindered by the fermion sign problem~\cite{signproblem1,signproblem2,signproblem3}. It should be noted that neutron stars,  possess not only a finite baryon density but also a finite isospin density, which arises due to isospin asymmetry. The study of QCD at finite isospin density (and zero baryon density)  is, however, unencumbered by the fermion sign problem~\cite{Son1, Son2, Son3, Kogut1, Kogut2, Kogut3}.  While this system, differs from the physical system of relevance in neutron stars, it does give a distinct handle to get insights into how QCD behaves in extreme environments.

Recently, finite isospin QCD in a background magnetic field and finite temperatures was studied on the lattice~\cite{Endrodi}. The study shows the existence of the fermion sign problem at finite isospin QCD with a magnetic field even though the sign problem is absent in finite isospin QCD when the magnetic field is absent. This occurs due to the breaking of flavor symmetry of the finite isospin system by the external magnetic field. Note that the electromagnetic charge of the up and down quarks are opposite in sign but different in magnitude unlike their chemical potentials, which are equal and opposite. The study circumvents the fermion sign problem by doing a Taylor expansion in the magnetic field. The results in that work suggest that at low temperatures, the system exhibits a uniform diamagnetic phase (i.e. negative magnetic susceptibility) for isospin chemical potentials that are larger than the pion mass but not asymptotically large.

In the present study, we study finite density isospin matter in a uniform magnetic field in a regime in which analytic calculations are legitimate.  We work at leading order in chiral perturbation theory ($\chi$PT) and consider the regime, where all physical parameters are much smaller than the typical hadronic scale ($\Lambda_{\rm Had} \sim 4 \pi f_{\pi}$). More specifically, the ratio
\begin{equation}
\frac{x}{\Lambda_{\rm Had}}\ll 1\ ,
\end{equation}
where $x$ can be the isospin chemical potential ($\mu_{\rm I}$), pion mass ($m_{\pi}$), pion momenta ($p$) or $\sqrt{eH}$, where $e$ is the charge of a pion and $H$ is the external magnetic field.   This ensures the validity of chiral perturbation theory.   Additionally, we assume that the baryon density is zero and only consider the effects of the isospin chemical potential. We show the existence not only of a uniform diamagnetic, superconducting state which persists at low magnetic fields but also show the existence of a topological, vortex phase.

A system of finite isospin density was first considered by Son and Stephanov in Refs.~\cite{Son1,Son2}, where it was seen that for isospin chemical potential ($\mu_{I}$) larger than  the pion mass ($m_{\pi}$), it is energetically favorable for pions to condense.  In that work, both weak and electromagnetic interactions were turned off. Here we continue to ignore pion decay via weak interactions; this is innocuous since the contributions of weak interactions are small.  The electromagnetic interactions among the pions, however are potentially  problematic: if a system with isospin density $n_{I}$ is confined to a region of spatial extent, $R$, then energy due to pion-pion electrostatic interactions scales as
\begin{equation}
E_{\rm electrostatic} \sim e^2 n_I^2 R^5  \; ,
\end{equation}
where $n_I$ is the isospin density which at these low densities is effectively the density of pions. The energy associated with strong interactions and the interaction with  the external magnetic field in the regime of validity of $\chi PT$ scales as
\begin{equation}
E_{\rm \chi PT} \sim f_\pi^4 \,  g \left ( \frac{m_\pi^2}{f_\pi^2}, \frac{n_I}{f_\pi^3}, \frac{e H}{f_\pi^2} \right ) R^3 \; ,
\end{equation}
where $g$ is a calculable dimensionless function of order unity.  Clearly in the thermodynamic limit of $R \rightarrow \infty$, $E_{\rm electrostatic}$ dominates.  Nevertheless, we will neglect it here even though we work in the thermodynamic limit. 

It is important to justify the neglect of the electrostatic interactions between pions.  Here, the calculations are not being used to directly describe a realistic physical situation but rather to gain some insight into how QCD behaves by working in a particular more tractable regime.  The issue is somewhat analogous to studies of infinite nuclear matter~\cite{Weisskopf}.  Clearly infinite nuclear matter has the same difficulty---the electrostatic energy per particle diverges.  Despite this, knowledge of the properties of infinite nuclear matter neglecting the electrostatic energy give important insights into nuclear physics even in regimes where the electrostatic energy cannot be neglected. 

For real world parameters, there are two regimes for which the approximation of simultaneously neglecting electromagnetic interactions between pions and working in the thermodynamic limit appears to be appropriate.  One is the regime in which the system is confined to a finite region satisfying
\begin{equation}
\frac{1}{f_\pi^2} \ll R^2 \ll \frac{1}{e^2 f_\pi^2} \; .
\end{equation}
The first condition is what is needed to justify the thermodynamic limit when other parameters are in the typical regime of validity of $\chi PT$; the second condition is what ensures that the electrostatic energies are small.  Note that in the Heaviside convention used here, $1/ e^2 \sim 137$ so that there exists an $R^2$ which is an order of magnitude bigger than $1/f_{\pi}^2$ while being an order of magnitude smaller than $ \frac{1}{e^2 f_\pi^2} $.  An alternative regime is a thermodynamically large one in which the positive electric charge of the isospin matter is neutralized by a background of electrons.  This situation is complicated by the fact that these ``background" electrons also couple to the external magnetic field and are known to exhibit superconducting properties which might affect the behavior of the pions.  However, the fact that electrons are so much lighter than the pions renders the electrons innocuous over the region of interest.
\ As shown in Refs.~\cite{Bailin1,Bailin2}, the zero-angular momentum, positive parity ($0^{+}$) pairing of relativistic electrons can exhibit either type-I or type-II superconductivity. Since electron masses are negligible compared to pion masses, electrons in our system move ultra-relativistically, where the behavior of the superconductor is strongly type-I. The magnetic field ($H^{e}_{c}$) at which the phase transition from a superconducting state to a normal state occurs is of $\mathcal{O}(k_{\rm F}k_{\rm B}T_{c})$ at zero temperature. Here $k_{\rm F}$ is the Fermi momenta, $k_{\rm B}$ is the Boltzmann constant (an in our units is just 1) and $T_{c}$ is the critical temperature below which superconductivity occurs. As will be shown later, the magnetic fields required for which pions undergo phase changes are much larger and of $\mathcal{O}(f_{\pi}^{2})$, where $f_{\pi}$ is the pion decay constant. Therefore, except for a narrow regime below $H^{e}_{c}$, the superconducting properties of the ultra-relativistic electrons do not interfere with that the pions.

\section{$\chi$PT Lagrangian at finite isospin}
We begin with a brief review of the finite isospin Lagrangian at leading order in $\chi$PT that was first considered by Son and Stephanov in Refs.~\cite{Son1, Son2}. The relevant regime where the Lagrangian is valid is when $\frac{x}{\Lambda_{\rm Had}}\ll 1$, where $x$ could be the pion momenta, pion mass ($m_{\pi}$), the isospin chemical potential $\mu_{I}$ or $\sqrt{eH}$, where $H$ is the external magnetic field and $e$ the pion charge. The effective Lagrangian has the form:
\begin{equation}
\label{Lagrangian}
\mathcal{L}_{\rm eff}=f_{\pi}^{2}\ \textrm{Tr} (D_{\mu}\Sigma^{\dagger}D^{\mu}\Sigma)+m_{\pi}^{2}f_{\pi}^{2}\textrm{Tr}(\Sigma+\Sigma^{\dagger}) \ ,
\end{equation}
where $\Sigma$ represents SU(2) matrices, $m_{\pi}$ the pion mass and $f_{\pi}$ the pion decay constant. The covariant derivatives are defined as follows: 
\begin{equation}
\label{covariant}
D_{\mu}\Sigma=\partial_{\mu}\Sigma-i\ [\delta_{\mu 0}\mu_{\rm I},\ \Sigma]
\end{equation}
with the isospin chemical potential entering the Lagrangian as the zeroth component.   
The values of the pion mass and pion decay constant are approximately $135\ MeV$ and $93\ MeV$ respectively. In order to proceed, we choose the following ansatz for the SU(2) matrix $\Sigma$:
\begin{equation}
\label{Sigma}
\begin{split}
\Sigma&=\frac{1}{f_{\pi}}\left( \sigma\mathbb{1}+i\pi_{x}\tau_{1}+i\pi_{y}\tau_{2}+i\pi_{z}\tau_{3} \right )\\
&=\cos\psi(\cos\theta\mathbb{1}+i\sin\theta(\cos\alpha\tau_{1}+\sin\alpha\tau_{2}))+i\sin\psi \tau_{3}
\end{split}
\end{equation}
Note that the neutral pion is represented by $\pi_{z}$ and the positively and negatively charged pions by $\pi_{x}\pm i\pi_{y}$ respectively.
Plugging this ansatz into Eq.~(\ref{Lagrangian}), the effective Lagrangian becomes:
\begin{widetext}
\begin{equation}
\mathcal{L}_{\rm eff}=-\frac{f_{\pi}^{2}}{2}\left [ \cos^{2}\psi \{\sin^{2}\theta(\vec{\nabla}\alpha )^{2} + (\vec{\nabla}\theta)^{2}\} +(\vec{\nabla}\psi)^{2}\right ]+m_{\pi}^{2}f_{\pi}^{2}(\cos\theta\cos\psi-1)+\frac{\mu_{I}^{2}f_{\pi}^{2}}{2}\sin^{2}\theta\cos^{2}\psi
\end{equation}
\end{widetext}
Note that the lagrangian has been normalized such that $\mathcal{L}_{\rm eff}=0$ when $\theta=0$, which is the normal QCD vacuum state at zero isospin or as shown in Refs.~\cite{Son1, Son2}, the vacuum at finite isospin for isospin chemical potentials less than or equal to the pion mass. 

From the Lagrangian density, it is straightforward to deduce the ground state of the system that was worked out in Refs.~\cite{Son1,Son2}. In order to so, we assume that the kinetic energy is zero and maximize the remaining Lagrangian. Since each of the potential energy contributions are of the same sign and is a function of $\cos\psi$, the ``energy" (more specifically $\mathcal{H}+\mu\ n$, where $\mathcal{H}$ is the Hamiltonian, $\mu$ is the chemical potential and $n$ is the number density) is minimized for $\psi=0$. Then the Lagrangian has to be maximized with respect to $\theta$, which leads to two possible solutions: either $\sin\theta=0$ or $\cos\theta=\frac{m_{\pi}^{2}}{\mu_{\rm I}^{2}}$. If $\sin\theta=0$, then $\mathcal{L}_{\rm eff}=0$ but if $\cos\theta=\frac{m_{\pi}^{2}}{\mu_{\rm I}^{2}}$, then $\mathcal{L}_{\rm eff}=\frac{f_{\pi}^2(\mu_{\rm I}^{2}-m_{\pi}^{2})^{2}}{2\mu_{\rm I}^{2}}$. Therefore, when the isospin chemical potential is greater than the pion mass, the condensed phase ($\theta\neq 0$) is energetically more favorable compared to the normal phase ($\theta=0$).  

\section{Gibbs Free Energy}
Here we want to study the effect of coupling the pions to dynamical photons and a uniform, external magnetic field. The effective Lagrangian for this particular system is easily obtained by introducing photon fields and changing the covariant derivation of Eq.~(\ref{covariant}) to include electromagnetic gauge fields. The new covariant derivative is as follows:
\begin{equation}
D_{\mu}\Sigma=\partial_{\mu}\Sigma-i\ [\delta_{\mu 0}\mu_{\rm I},\ \Sigma]-ieA_{\mu}[Q,\ \Sigma]\ ,
\end{equation}
with $Q$ being the charge matrix for the quarks and is defined as:
\begin{equation}
Q=\frac{1}{6}\mathbb{1}+\frac{1}{2}\tau_{3}\ ,
\end{equation}
where $\mathbb{1}$ is a $2\times 2$ identity matrix and $\tau_{3}$ is the third Pauli matrix. The resulting effective Lagrangian using the definition $\Sigma$ from Eq.~(\ref{Sigma}) is as follows:
\begin{widetext}
\begin{equation}
\label{Leff}
\mathcal{L}_{\rm eff}=-\frac{1}{4}F_{ij}F^{ij}-\frac{f_{\pi}^{2}}{2}\left [ \cos^{2}\psi \{\sin^{2}\theta\left (\vec{\nabla}\alpha+e\vec{A} \right)^{2} + (\vec{\nabla}\theta)^{2}\} +(\vec{\nabla}\psi)^{2}\right ]+m_{\pi}^{2}f_{\pi}^{2}(\cos\theta\cos\psi-1)+\frac{\mu_{I}^{2}f_{\pi}^{2}}{2}\sin^{2}\theta\cos^{2}\psi\ ,
\end{equation}
\end{widetext}
where $F_{ij}\equiv \partial_{i}A_{j}-\partial_{j}A_{i}$ is the electromagnetic tensor. We have assumed here that the zeroth component of the four-potential $A_{\mu}$ vanishes.

In order to consider the thermodynamics of the finite isospin system (with photons) coupled to a uniform background magnetic field, it is standard to consider the Gibbs free energy density, which is defined as~\footnote{Note that we are using Lorentz-Heaviside units where explicit factors of $4\pi$ do not appear.}:
\begin{equation}
\label{G}
\mathcal{G}=\mathcal{H}_{\rm eff}-\vec{M}\cdot\vec{H}\ .
\end{equation}
Here the magnetization $\vec{M}$ is defined as $\vec{M}\equiv\vec{B}-\vec{H}$, with $\vec{B}\equiv\vec{\nabla}\times\vec{A}$. As it is standard, we will assume that the external magnetic field, which we will label $\vec{H}$ only has a z-component. Since we are considering a time-independent system, the Hamiltonian density is given by the relation $\mathcal{H}_{\rm eff}=-\mathcal{L}_{\rm eff}$, where the $\mathcal{L}_{\rm eff}$ is defined in Eq.~(\ref{Leff}).

\section{Superconductivity}
At finite isospin, the condensed phase of pions is a superfluid with one of the charged pions forming a massless mode while the neutral pion and the other charged pion is massive~\cite{Son1,Son2}. It is natural to expect then that this superfluid phase exhibits superconducting behavior in the presence of an external magnetic field: the system sets up currents that produce opposing magnetic fields to cancel out the external magnetic field. For the finite isospin system in an external magnetic field, a natural question that arises then is the nature of superconductivity that the system exhibits, either type-I or type-II. The paradigm for studying superconductivity is Ginzburg-Landau theory or the Abelian Higgs model. However, the effective Lagrangian of Eq.~(\ref{Leff}) for the finite isospin system is neither isomorphic to Ginzburg-Landau~\cite{Landau,Abrikosov,Gorkov} or consequently the Abelian Higgs model. Also note that the finite isospin system consists of an additional degree of freedom (namely the neutral pion) in addition to the charged pions. Therefore, relative to Ginzburg-Landau, there is an extra degree of freedom in our problem. Furthermore, the effective Lagrangian for our system seems highly non-linear due to the presence of infinite number of non-linear terms present through the sine and cosine functions that characterize the pion fields through $\Sigma$ defined in Eq.~(\ref{Sigma}).

While the Gibbs free energy of our problem seems quite different from that of either Ginzburg-Landau theory or the Abelian Higgs model, it is still useful to borrow insights from the well known problems. It is well-known in these theories that the system exhibits type-I superconductivity if:
\begin{equation}
\xi>\sqrt{2}\lambda\ ,
\end{equation}
and type-II superconductivity if:
\begin{equation}
\xi<\sqrt{2}\lambda\ .
\end{equation}
Here $\xi$ is the coherent length of the macroscopic state, and $\lambda$ is the penetration depth of the external magnetic field. The superconductivity type is determined by considering the surface energy of the interface consisting of a half-infinite normal state and a half-infinite superconducting state at the critical magnetic field, where the normal and superconducting states have the same Gibbs free energy. If the surface energy is positive at the critical magnetic field, then the system makes a first order transition from the superconducting state to a normal. However, if the surface energy is negative, then an intermediate state of Abrikosov vortices are formed such that the transition to the normal state from the superconducting state becomes second order.

While the exact mathematical results of when superconductivity is type-I or type-II does not directly apply for the case of finite isospin QCD, it is still useful to compare the relative sizes of the coherence length and penetration depth in our problem. Since the coherent length and penetration depth are related to the masses of the pions and the photons, we proceed by expanding the effective Lagrangian of Eq.~(\ref{Leff}) about the ground state of the condensed, superfluid phase $\theta_{\rm gs}=\arccos\left ( \frac{m_{\pi}^{2}}{\mu_{\rm I}^{2}}\right )$ and $\psi_{\rm gs}=0$. In doing so, we find that the photon mass $m_{\rm A}$ is:
\begin{equation}
m_{\rm A}=e f_{\pi} \sin^{2}\theta_{\rm gs}=q f_{\pi} \left (1-\frac{m_{\pi}^{4}}{\mu_{I}^{4}}\right )\ ,
\end{equation}
and the mass of the charged pion is given by
\begin{equation}
\label{mtheta}
m_{\theta}=\sqrt{\left ( m_{\pi}^{2}\cos\theta_{\rm gs}-\mu_{\rm I}^{2}\cos2\theta_{\rm gs}\right )}=\mu_{\rm I}\sqrt{\left( 1-\frac{m_{\pi}^{4}}{\mu_{\rm I}^{4}}\right ) }\ .
\end{equation}
Note that here $e$ is the pion charge, $f_{\pi}$ is the pion decay constant and $m_{\pi}$ is the pion mass in the normal phase. We do not consider the mass of the neutral pion. As will be seen later, this is justified since the neutral pions do not condense.

Since the coherent length is inversely proportional to the pion mass, i.e. $\xi\sim\frac{1}{m_{\theta}}$ and the penetration depth is inversely proportional to the photon mass, i.e. $\lambda\sim\frac{1}{m_{\rm A}}$, it is useful to compare the masses of the pion and the photon to determine whether our system behaves as a type-I or type-II superconductor.

\begin{figure}[h]
    \centering
    \includegraphics[width=0.5\textwidth]{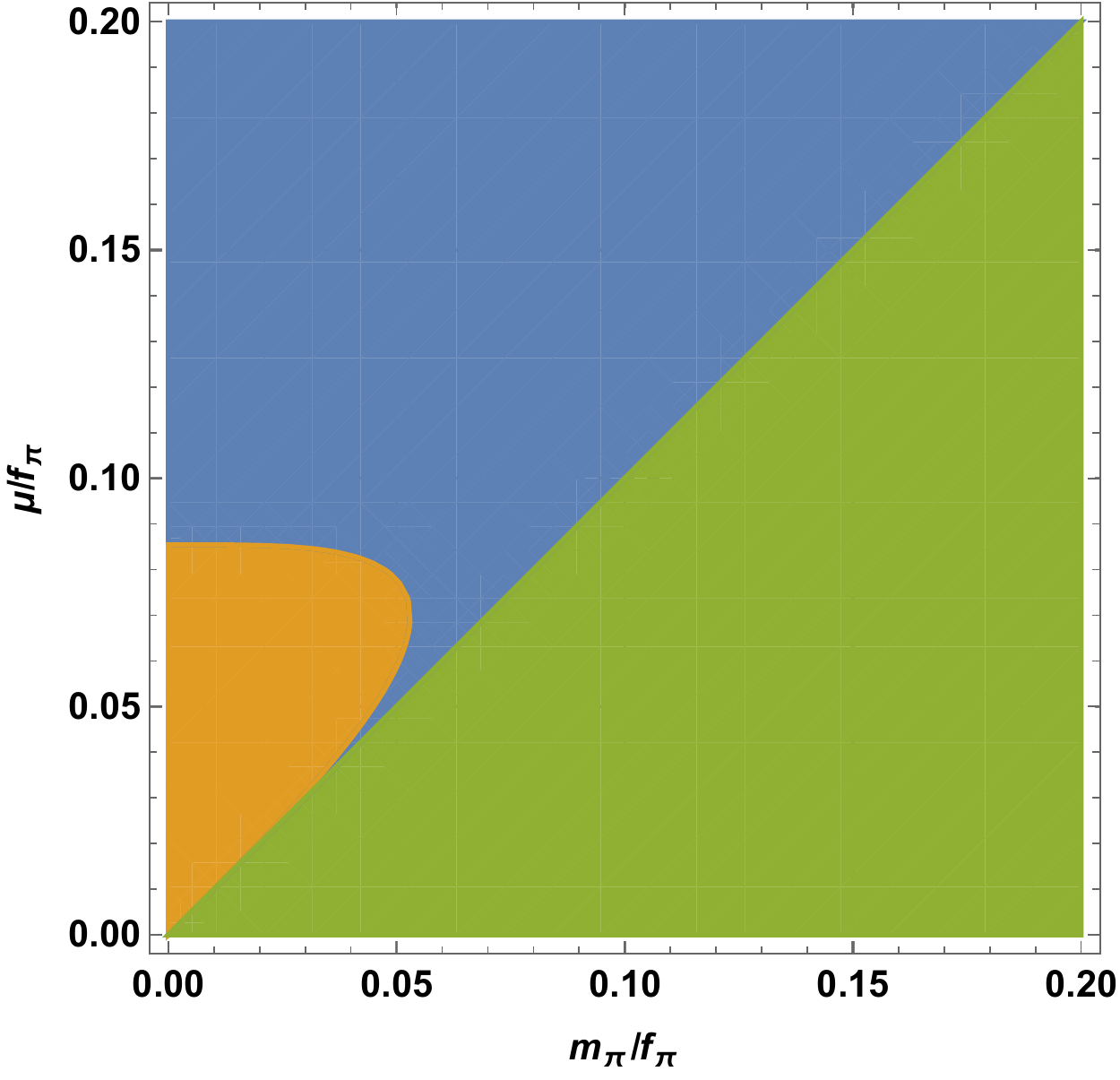}
    \caption{The region where $m_{\rm A}>m_{\theta}$ is shown in blue, the region where $m_{\theta}>m_{\rm A}$ is shown in orange and the green region represents the normal phase i.e. $\mu_{\rm I}\le m_{\pi}$. }
    \label{masscomparison}
\end{figure}

In Fig.~\ref{masscomparison} we have a plot showing a region where the mass of the photon is larger than the mass of the pion (in orange) and a region where the photon mass is smaller than the pion mass (in blue). The green region is where the normal phase exists and is therefore, irrelevant.

While the regions do not exactly delineate regions where the superconductivity is type-I or type-II, they do still prove some qualitative insight into the nature of superconductivity we might expect. For instance, it is reasonable to assert that in regime of the parameter space where $m_{A}\ll m_{\theta}$ should be a type-I superconductor and the regime where $ m_{\rm A}\gg m_{\theta}$ should be a type-II superconductor. Therefore, there is a narrow region at low isospin chemical potentials and small (but unrealistic) pion masses (near the origin) where the system will behave as a type-I (suggesting the existence of a critical point for small pion masses since the system goes from type-I to type-II with increasing isospin chemical potentials). However, for most of the parameter space, and in particular for realistic pion masses, the system will behave as a type-II superconductor. 

\subsection{Type-I}

Here, we briefly consider type-I superconductivity and determine the critical magnetic field at which the transition from the normal to the superconducting state occurs. First, we consider the Gibbs free energy density of the normal state, which occurs when $\theta=0$. In this case, the external field completely penetrates the vacuum state such that the magnetic field ($\vec{B}$) of the normal state is equal to the external magnetic field ($\vec{H}$). The Gibbs free energy of the system in its normal phase is
\begin{equation}
\mathcal{G}_{\rm n}=\mathcal{H}_{\rm n}-\vec{M}_{\rm n}\cdot \vec{H}=\frac{1}{2}\vec{H}^{2}\ .
\end{equation}
Here $\mathcal{H}_{\rm n}$ represents effective Hamiltonian in the normal phases and $M_{\rm n}$ is the magnetization in the normal phase. The second equality is obtained by using Eqs.~(\ref{G}) and (\ref{Leff}) and the fact that the phase is spatially homogeneous with $\theta=0,\ \psi=0$. 

However, in the condensed phase with $\theta\neq 0$, the magnetic field cannot penetrate the state. The photon fields readjust themselves such that the field entering the system is zero. i.e. $\vec{B}=0$. Therefore, the Gibbs free energy of the condensed phase, which is superconducting is:
\begin{equation}
\label{sc}
\mathcal{G}_{\rm s}=\mathcal{H}_{\rm s}-\vec{M}_{\rm s}\cdot\vec{H}=-\frac{f_{\pi}^2(\mu_{\rm I}^{2}-m_{\pi}^{2})^{2}}{2\mu_{\rm I}^{2}}+\vec{H}^{2}\ ,
\end{equation}
where $\mathcal{H}_{\rm s}$ is the Hamiltonian and $\vec{M}_{\rm s}$ is the magnetization of superconducting phase. In the superconducting ground state $\theta=\arccos\left (\frac{m_{\pi}^{2}}{\mu_{\rm I}^{2}} \right )$ and $\psi=0$.

The critical magnetic field at which the phase transition from the normal state to the superconducting state occurs when the Gibbs free energies of the normal state equals that of the superconducting state, i.e. $\mathcal{G}_{\rm n}=\mathcal{G}_{\rm s}$. This gives a critical field $\vec{H}_{c}$, which has the following magnitude
\begin{equation}
\label{Hc}
|\vec{H}_{c}|=\frac{f_{\pi}(\mu_{\rm I}^{2}-m_{\pi}^{2})}{\mu_{\rm I}}\ .
\end{equation}
It is not surprising that the size of the critical field increases with increasing isospin chemical potentials. The density of pions that condense increases with increasing chemical potentials, which in turn means that larger currents are generated and the superconducting state persists at larger external magnetic fields.

\subsection{Type-II} 
Next, we consider type-II superconductivity, which as discussed previously is expected to occur for realistic pion masses. Here, we will determine the first critical magnetic field $H_{c1}$ at which the transition from the uniform superconducting state to the vortex state occurs. We will also give a theoretical estimate of the second critical magnetic field $H_{c2}$ where the transition to the normal state occurs. However, we do not calculate explicitly the surface energy, which determines whether the system actually behaves as a type-II superconductor. Alternatively, we can just compare the first critical magnetic field ($H_{c1}$) with the critical magnetic field of Eq.~(\ref{Hc}) to determine that the system actually is type-II. If the first critical field $H_{c1}$ is smaller than the critical field $H_{c}$ of the type-I, then the Gibbs free energy of the vortex state is lower than the normal state and therefore, the system is type-II.

The intermediate vortex state is a cylindrically symmetric state. In order to determine the first critical point, we consider a single vortex state with the following ansatz:
\begin{equation}
\label{alpha}
\alpha=-n\phi\ ,
\end{equation}
where $\phi$ is the polar angle in cylindrical coordinates, which goes from $0$ to $2\pi$ and $n$ is an integer. It determines the amount of flux that is quantized within the vortex. This flux ($\Phi$) is given by the relation:
\begin{equation}
\label{Phi}
\Phi=\int_{\textrm{C}_{\infty}}\vec{A}\cdot d\vec{l}=\int_{\textrm{C}_{\infty}} \vec{B}\cdot d\vec{A}=\frac{2n\pi}{e}\ ,
\end{equation}
where the path integral is performed on the boundary of some region $\textrm{C}_{\infty}$, enclosing the vortex. The second equality arises through the use of the definition $\vec{B}\equiv\vec{\nabla}\times \vec{A}$ and Stoke's theorem. Note that the $n$ that appeared in Eq.~(\ref{alpha}) also appears here.

We can now write down the Gibbs free energy of a single vortex state with the lowest amount of flux going through it, i.e. $n=1$. We do so using the following choice of gauge fields:
\begin{equation}
\vec{A}=\left(A_{\rm r},\ A_{\phi},\ A_{\rm z}\right )=\left ( 0,\ A_{\phi}(r),\ 0\right )\ .
\end{equation}
Again, we have defined the gauge choice in cylindrical coordinates owing to the cylindrical symmetry of the vortex state. The physical implication of the vortex ansatz is easy to understand from the definition of the electromagnetic current, which is easily determined through the effective Lagrangian of Eq.~(\ref{Leff}):
\begin{equation}
\label{j}
\vec{j}=\frac{\partial\mathcal{L}}{\partial \vec{A}}=q\cos^{2}\psi\sin\theta\left ( \vec{\nabla}\alpha+e\vec{A}\right)\ .
\end{equation}
Noting that the flux quantization condition implies that:
\begin{equation}
\lim_{r\rightarrow\infty}A_{\phi}(r)=\frac{1}{er}\ ,
\end{equation}
which along with the definition of the electromagnetic current in Eq.~(\ref{j}), implies that the current at the boundaries of the vortex vanishes.

Using the flux quantization condition then, the Gibbs free energy of a single vortex state with $n=1$ is:
\begin{widetext}
\begin{equation}
\begin{split}
\label{Gvortex}
\mathcal{G}_{\rm vortex}&=\mathcal{G}_{\rm t}-\mathcal{G}_{\rm s}\\
\mathcal{G}_{\rm t}&=\frac{1}{2}\left( \frac{1}{r}\frac{\partial(rA_{\phi})}{\partial r} \right )^{2}+\frac{f_{\pi}^{2}}{2}\left [ \cos^{2}\psi \{\sin^{2}\theta\left (-\frac{1}{r}+e A_{\phi} \right)^{2}+\left (\frac{\partial\theta}{\partial r}\right )^{2} \}+\left(\frac{\partial\psi}{\partial r}\right)^{2}\right ]\\
&-m_{\pi}^{2}f_{\pi}^{2}\cos\theta\cos\psi-\frac{\mu_{I}^{2}f_{\pi}^{2}}{2}\sin^{2}\theta\cos^{2}\psi-\vec{M}_{\rm t}\cdot\vec{H}\\
\vec{M}_{\rm t}&=(\vec{B}_{\rm t}-\vec{H})\\
\vec{B}_{\rm t}&=\hat{\mathbf{z}}\left( \frac{1}{r}\frac{\partial(rA_{\phi})}{\partial r} \right )\ ,
\end{split}
\end{equation}
\end{widetext}
where $\mathcal{G}_{\rm s}$ is the Gibbs free energy of the uniform superconducting state, which was calculated in Eq.~(\ref{sc}).

Since it is not possible to find the solutions of the vortex analytically by solving for the gauge field $\vec{A}$ and the pion field implicitly represented by $\theta$ and $\alpha$, we will proceed to numerically find the solution of the vortex state. An important question arises in the context of solving for the vortex state. It pertains to the fact that in chiral perturbation theory, in addition to the charged pions there are also neutral pions. These neutral pions are uncondensed as was shown in Section II based on results of Refs.~\cite{Son1, Son2}. It is important to establish that that this remains the case in finite magnetic fields.

Formally, checking that the neutral pion fields indeed do not condense i.e. $\psi=0$, amounts to finding solutions for charged pions under the assumption that the neutral pions do not condense and checking that the solutions are stable against fluctuations of the pion field. This can be done using the equation of motion for $\psi$ fields. Assuming radial symmetry for this field, the equation of motion is as follows:
\begin{widetext}
\begin{equation}
\partial_{\rm t}^{2}\psi(r)-\vec{\nabla}^{2}\psi(r)=f_{\pi}^{2}\left (2\cos\psi\sin\psi{\sin^{2}\theta\left (-\frac{1}{r}+eA_{\phi} \right )} \right )-m_{\pi}^{2}f_{\pi}^{2}\cos\theta\sin\psi-\mu_{\rm I}^{2}f_{\pi}^{2}\sin^{2}\theta\cos\psi\sin\psi\ .
\end{equation}
\end{widetext}
Expanding $\psi$ around its ground state value, i.e. $\psi_{\rm g.s.}=0$, such that $\psi=\psi_{\rm g.s.}+\delta\psi$ and assuming $\delta\psi={\rm Re}\sum_{n}\exp{(iE_{n}t)}\delta\tilde{\psi}_{n}(r)$, we obtain the following equation:
\begin{widetext}
\begin{equation}
\begin{split}
&-\vec{\nabla}^{2}\delta\tilde{\psi}_{n}+f_{\rm v}\delta\tilde{\psi}_{n}=E_{n}^{2}\delta\tilde{\psi}_{n}\\
&f_{\rm v}=\left ( \sin^{2}\theta(r)\left(-\frac{1}{r}+eA_{\phi}(r)\right)^{2}+\left(\frac{\partial\theta(r)}{\partial r} \right )^{2}\right )-m_{\pi}^{2}f_{\pi}^{2}\cos\theta(r)-\mu_{\rm I}^{2}f_{\pi}^{2}\sin^{2}\theta(r)\ ,
\end{split}
\end{equation}
\end{widetext}
where $f_{\rm v}$ is obtained by plugging the vortex solutions found by assuming that the neutral pions do not condense. If none of the the eigenvalues of the above equation are negative, i.e they satisfy $E_{n}^{2}\ge 0$ for all $n$, the vortex is stable against neutral pion condensation; if $E_{n}^{2}<0$ for some $n$, then the vortex is unstable. However, it is difficult to explicitly determine the sign of even just the smallest valued $E_{n}^{2}$ since we can only numerically solve for the vortex solution, which in turn affects the exact nature of $f_{\rm v}$. Therefore, we will proceed by minimizing the free energy of the vortex state with respect to all the degrees of freedom including the gauge field $A$ and the pion fields, $\theta$ and $\psi$. In doing so we find that it is energetically favorable for the neutral pion to remain uncondensed.

\subsection{Numerical Results}
In this subsection, we solve for the vortex solutions of the system directly by minimizing the Gibbs free energy of Eq.~(\ref{Gvortex}) and determine the first critical field. For the purposes of this numerical work, we will use the pion decay constant, $f_{\pi}$, to set the scale in the problem. We will introduce the following change of variables:
\begin{equation}
\begin{split}
\tilde{r}&=f_{\pi} r\\
\tilde{m}_{\pi}&=\frac{m_{\pi}}{f_{\pi}}\\
\tilde{\mu}_{\rm I}&=\frac{\mu_{\rm I}}{f_{\pi}}\\
\tilde{A}_{\phi}&=\frac{A_{\phi}}{f_{\pi}}\ .
\end{split}
\end{equation}

\begin{figure}[h]
    \centering
    \includegraphics[width=0.5\textwidth]{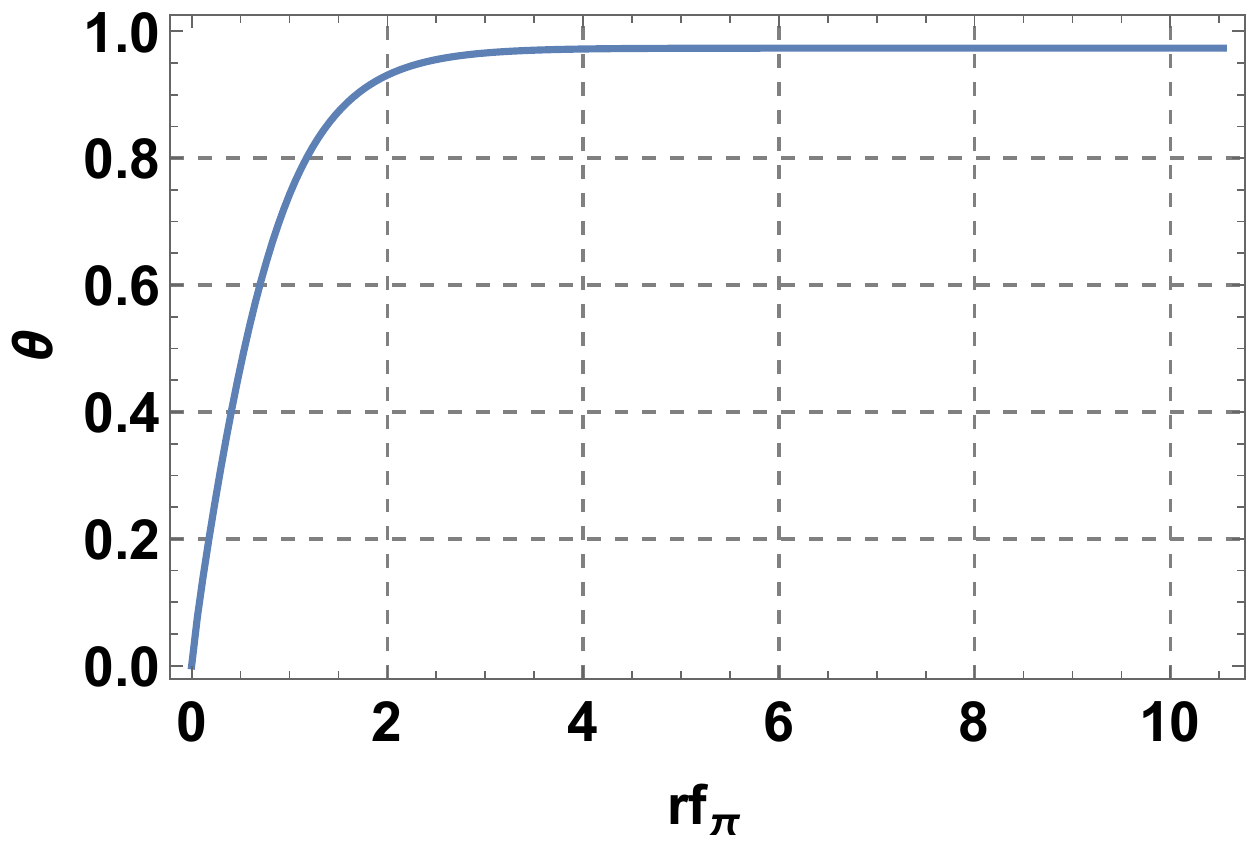}
    \caption{The plot shows the density of the charged pion as a function of radial distance from the center of the vortex, which is at the origin. The result is for $\tilde{m}_{\pi}=1.5$ and $\tilde{\mu}_{I}=2.0$.}
    \label{thetaplot}
\end{figure}

\begin{figure}[h]
    \centering
    \includegraphics[width=0.5\textwidth]{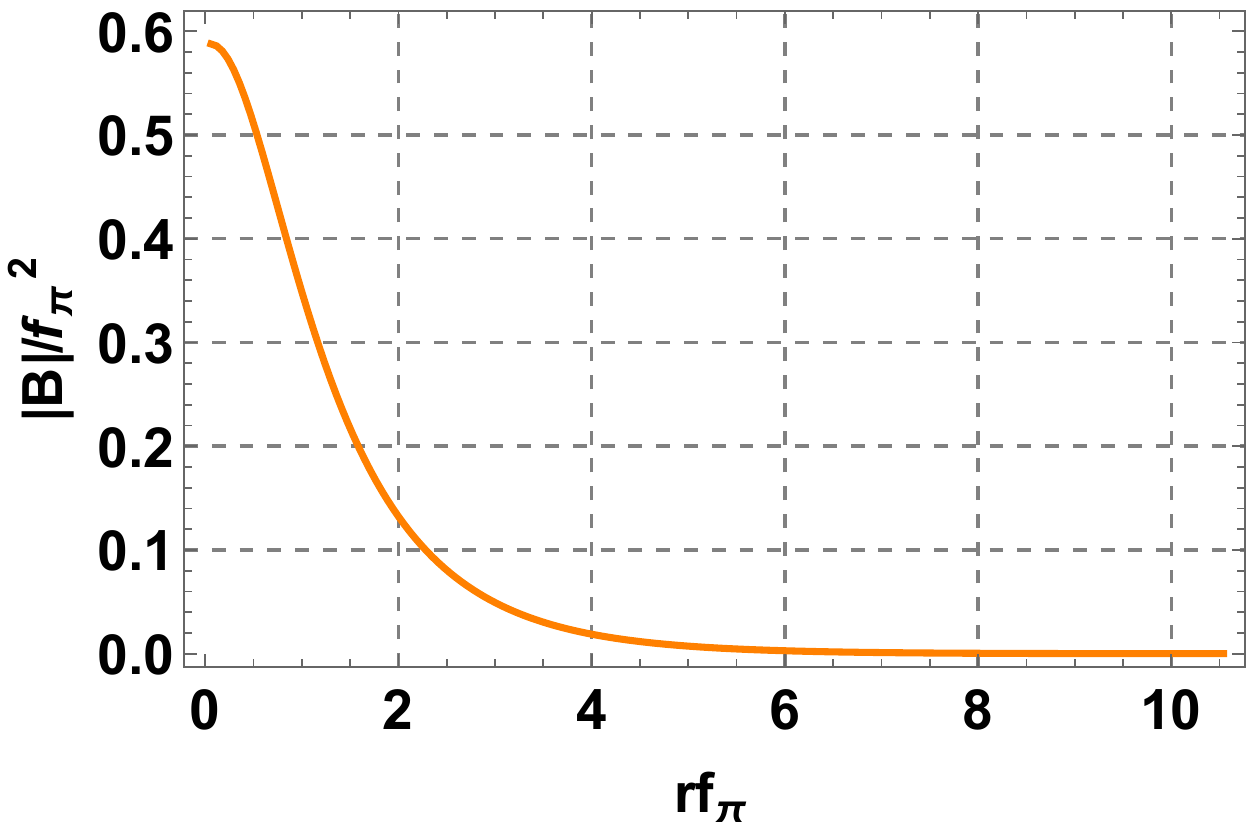}
    \caption{The plot shows the magnetic field i.e. $\vec{B}\equiv\vec{\nabla}\times\vec{A}$ as a function of radial distance from the center of the vortex. The result is for $\tilde{m}_{\pi}=1.5$ and $\tilde{\mu}_{I}=2.0$.}
    \label{Bplot}
\end{figure}

The numerical work is performed by discretizing the Hamiltonian and minimizing the energy per unit length, which is obtained from the discretized Hamiltonian through a numerical integration. The numerical error is dominated by the discretization of the Hamiltonian and the error that arises in the energy per unit length is of $\mathcal{O}\left( a^{2} \right)$, where $a$ is the size of each cell. The error arising from the numerical integration itself is $\mathcal{O}\left ( \frac{\Delta r^{3}}{N^{2}}\right)$, where $N$ is the number of discretized points and $\Delta \tilde{r}$ is the length of the integration region. Therefore, the numerical estimation of the critical magnetic field of Eq.~(\ref{Hc1}), in particular $\tilde{H}_{c1}=\frac{H_{c1}}{f_{\pi}^{2}}$ has an error approximately of $\mathcal{O}(a^{2})$.  For our calculation the maximum value of $\Delta {r}$ is 22 and $N$ was chosen to be 160. Therefore, the numerical errors in the results we present are relatively small.

For realistic pions masses, $\tilde{m}_{\pi}=1.5$ and we solve for single vortex solutions up to an isospin chemical potential of $\tilde{\mu}_{\rm I}=2.5$. These values are well within the regime of validity of chiral perturbation theory, which was discussed earlier. In Figs.~\ref{thetaplot} and \ref{Bplot}, we plot the radial profile of a single vortex solution. We omit the plot for $\psi(r)$ since it remains uniformly zero.

In generating the solutions, we assumed that $e=1$. Note that at the center of the vortex both $\theta(r)$ and $A_{\phi}(r)$ must vanish to prevent a singularity from occurring at the origin. (This can be easily understood from the equations of motion written in cylindrical coordinates.) Additionally, far away from the vortex $\theta$ assumes its ground state value for a uniform superconducting state and the magnetic field $\vec{B}\equiv\vec{\nabla}\times\vec{A}$ also vanishes. 

\begin{figure}[h]
    \centering
    \includegraphics[width=0.5\textwidth]{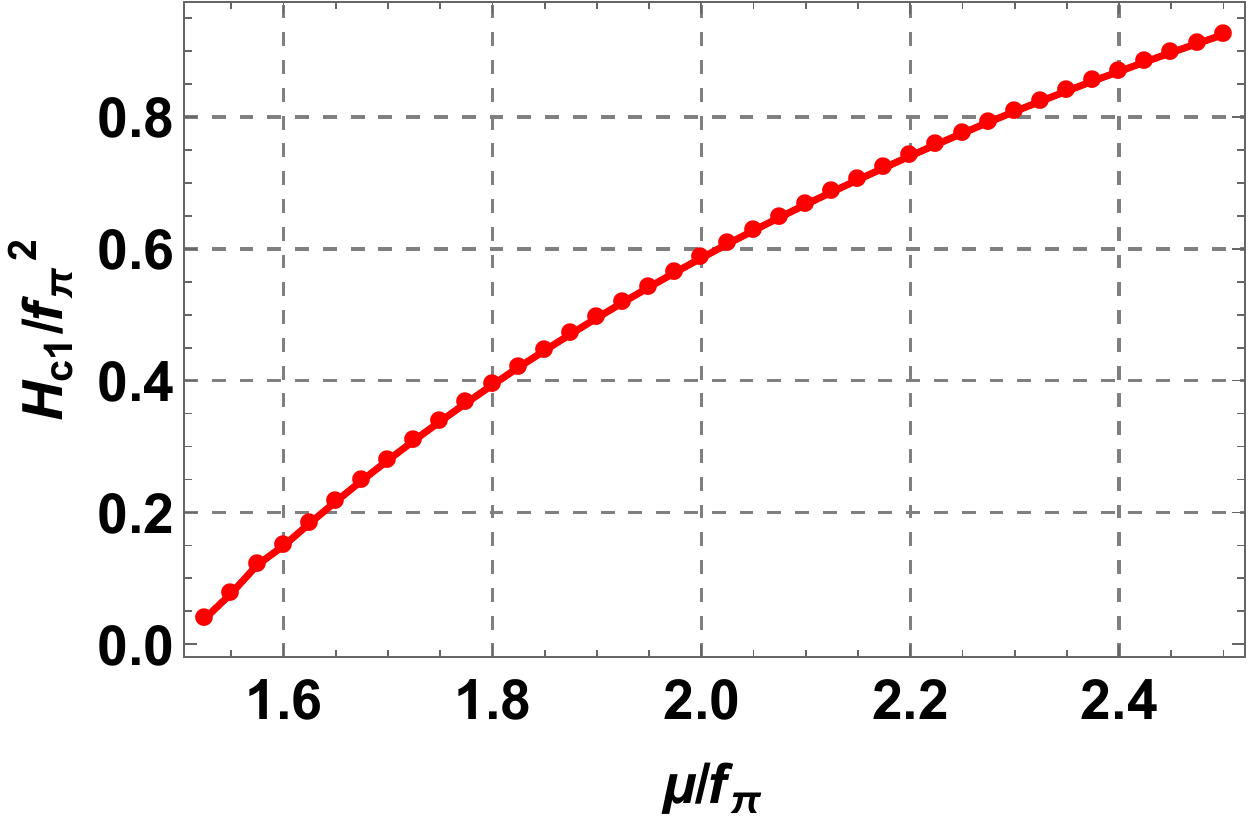}
    \caption{The plot shows the critical magnetic field for the formation of the vortex state for realistic pion mass: $\tilde{m}_{\pi}=1.5$ as a function of isospin density.}
    \label{criticalHc1plot}
\end{figure}

We also determine numerically the first critical magnetic field ($H_{c1}$) at which the transition from a uniform superconducting state, which exists at low magnetic fields, to a state with a single vortex. It is a standard result~\cite{Tinkham, Fetter} that the critical field is given by the relation:
\begin{equation}
\label{Hc1}
H_{c1}=\frac{E_{\rm vortex}/L}{\Phi_{0}}\ ,
\end{equation}
where $E_{\rm vortex}/L$ is the energy per unit length of the vortex state, which is determined by integrating the Hamiltonian of the vortex over the cross section of the vortex, and $\Phi_{0}$ is a single quanta of flux passing through the vortex. The size of this flux is $\frac{2\pi}{e}$ with $e$ being the charge of the pion. The relation itself is easily determined by using the fact that at the critical field $H_{c1}$, the Gibbs free energy of the superconducting state is equal to the Gibbs free energy of the single vortex state, i.e. $\mathcal{G}_{\rm s}=\mathcal{G}_{\rm vortex}$.

The first critical magnetic field is presented in Fig.~\ref{criticalHc1plot}. The size of critical field increases with increasing isospin chemical potential. This is expected since larger isospin chemical potential results in a larger density of condensed pions in the condensed phase, which further means that larger currents to oppose the external magnetic field can be generated at the boundaries.

\subsection{Second critical field}
While it is possible to numerically determine the lower critical magnetic field, where the transition from a uniform superconducting state to a vortex state occurs, it is generally harder to determine the upper critical point from which the transition from a vortex state to a normal state occurs. As the magnetic field increases past the first critical point, the density of vortices also steadily increase forming Abrikosov lattices. Close to the upper critical point, the cores of the vortices (the size of which is determined by the charged pion mass $m_{\theta}$) begin to overlap such that the average spacing between the vortices of $\mathcal{O}(m_{\theta}^{-1})$. Each vortex, however, continues to carry a single quanta of flux $\Phi_{0}\equiv\frac{2\pi}{e}$. Then, using Eq.~(\ref{Phi}), an estimate for the second critical point is given by
\begin{equation}
H_{c2}\sim\Phi_{0}m_{\theta}^{2}\ ,
\end{equation}
where $m_{\theta}$ is given in Eq.~(\ref{mtheta}).

\section{Final Comments}
In this paper, we have investigated finite isospin QCD in an external magnetic field at lowest order in chiral perturbation theory. We have shown here that the system forms a type-II superconductor for realistic pion masses and possibly also a type-I superconductor for small pion masses. It will be interesting to see if the vortex phases that have been observed here are also seen in lattice calculations. It is noteworthy that lattice simulations of finite isospin system in a magnetic field have already been carried out~\cite{Endrodi}. We note however, that the setup of this calculation apparently excluded  the physics associated with type-II superconductivity, for which vortices form and thereby alter the magnetic fields, i.e. the back-reaction of the pion field dynamics on the $B$ field appears to be absent.

%

\begin{acknowledgements}
P.A. and T.D.C. acknowledge the support of the U.S. Department of Energy through grant number DEFG02-93ER-40762. P.A. would also like to thank Shmuel Nussinov for some very illuminating discussions. J.S. would like to acknowledge the support of the Maryland Center of Fundamental Physics at the University of Maryland for their support during the course of this work.
\end{acknowledgements}

\end{document}